\documentclass[aps,prb,reprint,superscriptaddress,showkeys,footinbib,floatfix]{revtex4-2}

\usepackage[T1]{fontenc}				
\usepackage[utf8]{inputenc}			
\usepackage[english]{babel}			
\usepackage[protrusion=false]{microtype}	
\usepackage{upgreek}

\usepackage{color} 						
\definecolor{red}{rgb}{1,0,0}				
\definecolor{blue}{rgb}{0,0,1}				
\definecolor{black}{rgb}{0,0,0}				

\usepackage{soul} 
\definecolor{hlyellow}{rgb}{0.95,0.95,0}
\definecolor{hlgreen}{rgb}{0,0.95,0}
\sethlcolor{hlyellow}

\soulregister \ref 7
\soulregister \cite 7
\soulregister \citet 7
\soulregister \footnote 8
\soulregister {\ } 7
\soulregister \figref 7
\soulregister \tabref 7
\soulregister \secref 7
\soulregister \refref 7
\soulregister \eqref 7

\usepackage{amsmath} 
\usepackage{amssymb,amsfonts,mathrsfs} 
\usepackage{array} 
\DeclareMathOperator{\Tr}{Tr}

\usepackage{graphicx} 

\usepackage{hyperref} 
\definecolor{dullmagenta}{rgb}{0.4,0,0.4} 
\definecolor{darkblue}{rgb}{0,0,0.4}
\definecolor{medblue}{rgb}{0,0,0.6}
\definecolor{lightblue}{rgb}{0,0,0.8}
\hypersetup{
	colorlinks=true, 		
	breaklinks=true,		
	linkcolor=lightblue,
	citecolor=lightblue,
	urlcolor=lightblue,
	filecolor=dullmagenta
}
\usepackage[all]{hypcap} 

\usepackage{fixgreek}
\usepackage{physmaths}
\usepackage{siunitx}

\newcommand{\etal}{\textit{et~al.~}}

\newcommand{\figref}[1]{Fig.~\ref{#1}} 
\newcommand{\tabref}[1]{Tab.~\ref{#1}} 
\newcommand{\secref}[1]{Sec.~\ref{#1}} 
\newcommand{\refref}[1]{Ref.~\cite{#1}} 
\newcommand{\eqnref}[1]{Eq.~\eqref{#1}} 

\newcommand{\supmat}{Supplemental Material} 

\begin{document}

\newcommand{\mytitle}
{Probing the Jaynes-Cummings Ladder with Spin Circuit Quantum Electrodynamics
}
\title{\mytitle}

\author{Tobias \surname{Bonsen}}
\affiliation{QuTech and Kavli Institute of Nanoscience, Delft University of Technology, 2628 CJ Delft, Netherlands}

\author{Patrick \surname{Harvey-Collard}}
\affiliation{QuTech and Kavli Institute of Nanoscience, Delft University of Technology, 2628 CJ Delft, Netherlands}

\author{Maximilian \surname{Russ}}
\affiliation{QuTech and Kavli Institute of Nanoscience, Delft University of Technology, 2628 CJ Delft, Netherlands}

\author{Jurgen \surname{Dijkema}}
\affiliation{QuTech and Kavli Institute of Nanoscience, Delft University of Technology, 2628 CJ Delft, Netherlands}

\author{Amir \surname{Sammak}}
\affiliation{QuTech and Netherlands Organization for Applied Scientific Research (TNO), 2628 CJ Delft, Netherlands}

\author{Giordano \surname{Scappucci}}
\affiliation{QuTech and Kavli Institute of Nanoscience, Delft University of Technology, 2628 CJ Delft, Netherlands}

\author{Lieven~M.~K. \surname{Vandersypen}}
\email[Correspondence to: ]{L.M.K.Vandersypen@tudelft.nl}
\affiliation{QuTech and Kavli Institute of Nanoscience, Delft University of Technology, 2628 CJ Delft, Netherlands}

\date{April 25, 2023}

\begin{abstract}
We report observations of transitions between excited states in the Jaynes-Cummings ladder of circuit quantum electrodynamics with electron spins (spin circuit QED).
We show that unexplained features in recent experimental work correspond to such transitions and present an input-output framework that includes these effects.
In new experiments, we first reproduce previous observations and then reveal both excited-state transitions and multiphoton transitions by increasing the probe power and using two-tone spectroscopy.
This ability to probe the Jaynes-Cummings ladder is enabled by improvements in the coupling-to-decoherence ratio, and shows an increase in the maturity of spin circuit QED as an interesting platform for studying quantum phenomena.
\end{abstract}

\maketitle

Spin qubits in gate-defined silicon quantum dots (QDs) are a promising platform for quantum computing thanks to their small footprint, excellent coherence properties, and compatibility with today's highly advanced semiconductor industry \cite{loss1998quantum,hanson2007spins,zwanenburg2013silicon,zwerver2022qubits}.
Circuit quantum electrodynamics with spins, or spin circuit QED for short, focuses on the coherent coupling of spin qubits to photons in high-quality-factor superconducting resonators.
This can be used to achieve long-range two-qubit gates and readout of the qubit state \cite{burkard2020superconductor}, paving the way to a scalable architecture for quantum computing based on spins in linked quantum-dot arrays \cite{vandersypen2017interfacing}.
Following advances of circuit QED with superconducting qubits \cite{blais2004cavity,wallraff2004strong,blais2020quantum}, spin circuit QED has been achieved in several device architectures by leveraging spin-charge hybridization to couple the electron spin to the resonator electric field \cite{trif2008spin,cottet2010spin,hu2012strong,srinivasa2016entangling}.
Experiments with single electron spins in silicon \cite{samkharadze2018strong,mi2018coherent} and multispin qubits in gallium arsenide \cite{landig2018coherent} have achieved spin-photon coupling strengths that exceed the resonator and qubit linewidths, thereby reaching the strong coupling regime. 
Subsequently, simultaneous resonant interaction between a resonator and two spins has been achieved \cite{borjans2020resonant}, followed by resonator-mediated interaction between two remote spins in the dispersive regime \cite{harvey2022coherent}.
Additionally, spin circuit QED has been employed to achieve spin-transmon coupling \cite{landig2019virtual} and single-shot gate-based readout of spin qubits \cite{zheng2019rapid}.

\begin{figure*}[t]
   \centering
   \includegraphics{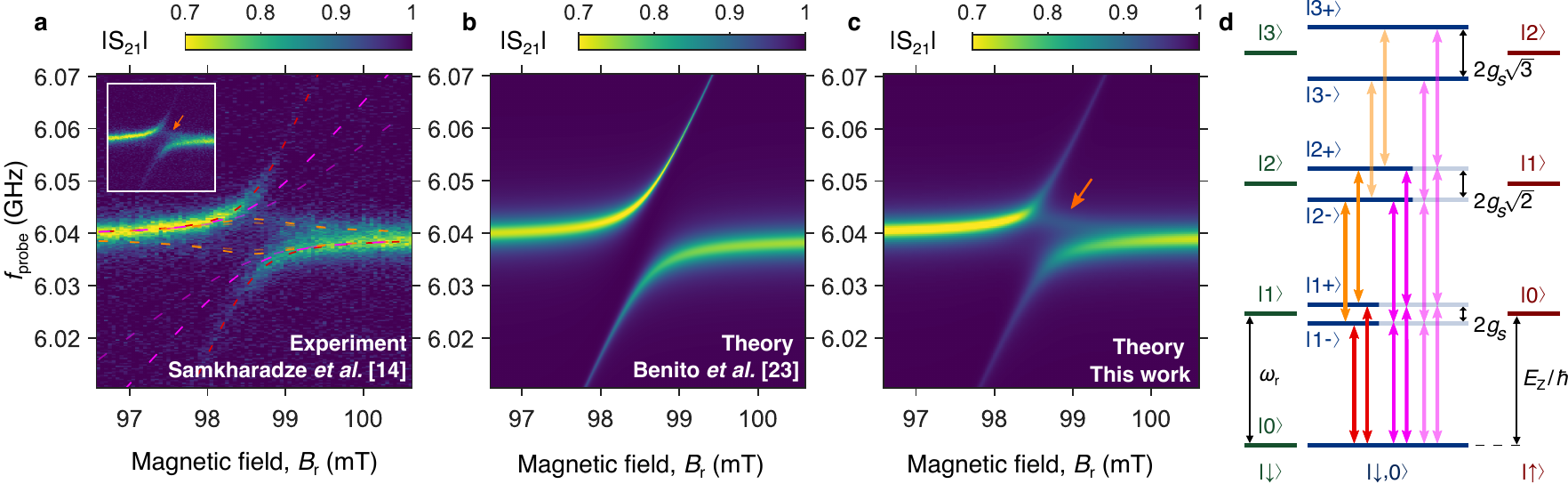}
   \caption{(a) Experimental resonator transmission spectrum from Samkharadze \etal\cite{samkharadze2018strong} for $2t_c/h=\SI{10.4}{\giga\hertz}$, together with transition frequencies (dashed lines) in the Jaynes-Cummings ladder in (d). Inset: the data is replotted without transition frequencies overlay. The avoided crossing demonstrates strong spin-photon coupling. An additional, previously unexplained feature appears inside the vacuum Rabi split peaks (arrow).
    (b) Transmission spectrum predicted by the standard input-output theory for spin circuit QED \cite{benito2017input} using the parameters in Supplemental Table~\ref{suptable:simparameters}. (c) Simulated spectrum using the theory presented in this work for probe amplitude $a_{\text{in,1}} =\SI{1000}{\hertz}^{1/2}$, thermal bath temperature $T=\SI{200}{\milli\kelvin}$, and other parameters in Supplemental Table~\ref{suptable:simparameters}. Since this experiment uses a hanger-style resonator, resulting in a resonance dip, the color scale has been inverted to match the transmission-style resonator data presented later in this work. (d) Transitions in the resonant spin-photon Jaynes-Cummings ladder:  main branches of the avoided crossing (red), excited-state transitions (orange) that correspond to the observed additional feature within the gap in (a), and multiphoton transitions (purple).}
    \label{fig1:Samkh2018_comparison}
\end{figure*}

The present work is motivated by results from the strong spin-photon coupling experiment of Samkharadze \etal\cite{samkharadze2018strong}. 
In this experiment, spin-charge hybridization was achieved by engineering an artificial spin-orbit interaction in a Si/SiGe double quantum dot (DQD).
The resulting spin-photon coupling was characterized with a spectroscopic measurement of the resonator transmission.
Specifically, the transmission as a function of probe frequency and magnetic field strength, reproduced here in \figref{fig1:Samkh2018_comparison}(a), shows a vacuum Rabi splitting of the modes, signaling the coherent hybridization of the spin with a single microwave photon.
An additional feature appears in the gap near the lower branch (arrow), which has remained unexplained until now. Additional peaks in a spectrum generally hint at the involvement of additional transitions in the system, which can spoil the behavior of resonator-mediated interactions. The development of a scalable spin-circuit-QED architecture will therefore require a deeper understanding of this phenomenon.

In this work, we explain the physical origin of the observed feature. We first find that its frequency matches transitions between excited states in the Jaynes-Cummings ladder.
Analogous signatures have also been observed in earlier circuit QED experiments with superconducting transmon qubits \cite{fink2008climbing,fink2010quantum}; however, in the spin-photon system they exhibit a different characteristic shape due to specific differences, and had not been identified as such.
We then develop a theoretical framework that combines input-output theory \cite{benito2017input, benito2020hybrid,kohler2018dispersive} with a Lindblad master equation \cite{benito2019optimized,manzano2020short}.
This theory captures transitions between excited states in the Jaynes-Cummings ladder, probe-power-dependent effects, and two-tone spectroscopy.
The simulated spectra reproduce well the observed feature in the vacuum Rabi splitting.
We show data from new experiments in which we both reproduce the observations of Samkharadze \etal~\cite{samkharadze2018strong} and furthermore reveal new multiphoton transitions~\cite{bishop2009nonlinear}.
We demonstrate the capability to drive some of these transitions, which could be useful for future photon preparation and detection schemes \cite{johnson2010quantum,albert2018performance}.

\begin{figure}[t]
    \centering
    \includegraphics{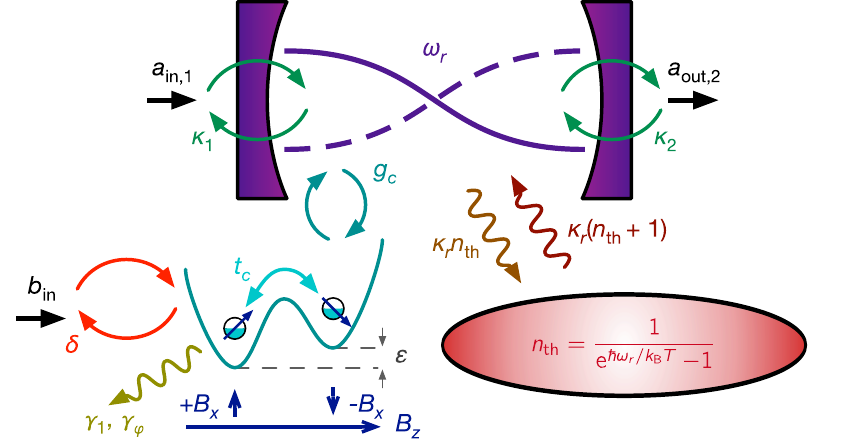}
    \caption{Overview of the input-output model for the coupled DQD-resonator system (see main text).}
    \label{fig2:schematic_model}
\end{figure}

The first step to explain the presence of the additional feature in the spectrum in \figref{fig1:Samkh2018_comparison}(a) is to identify the transitions involved.
To this end, we compare the data to the transition frequencies calculated from the system Hamiltonian $H=H_0+H_r+H_I$, see \figref{fig2:schematic_model}.
The full details of the spin-photon interaction have been described elsewhere \cite{hu2012strong,burkard2020superconductor,beaudoin2016coupling}.
The Hamiltonian for the double quantum dot containing one electron is given by 
\begin{equation} \label{eq:DQDHamiltonian}
    H_0 = \frac{1}{2}(\epsilon\tau_z+2t_c\tau_x+g_e\mu_B B_z\sigma_z+g_e\mu_B B_x\sigma_x\tau_z),
\end{equation}
where $\tau_\alpha$ and $\sigma_\alpha$ are the Pauli operators for position (left, right) and spin ($\uparrow,\downarrow$), respectively, $g_e=2$ is the Landé $g$ factor in silicon and $\mu_B$ is the Bohr magneton. 
At zero charge detuning, i.e., when $\epsilon = \SI{0}{\micro\electronvolt}$, the electron charge eigenstates with energy splitting $2t_c$ (``charge qubit'') develop a significant charge dipole that enables charge-photon and spin-photon interaction. 
This interaction can be turned off by localizing the electron onto a single dot, i.e., $\abs{\epsilon} \gg \abs{t_c}$.
The applied external magnetic field (with magnitude $B_r$ in the experiments), together with micromagnets fabricated on top of the DQD gate structure, result in a magnetic field gradient at the location of the DQD.
The homogeneous magnetic field component $B_z$ induces most of the Zeeman splitting of the electron spin states and is related to $B_r$ using the micromagnet model in \supmat\ Sec.~\ref{supsec:micromagnetmodel}, while the interdot magnetic field difference $2B_x$  causes spin and orbital states to hybridize \cite{beaudoin2016coupling}.
The resonator is modeled as a single-mode harmonic oscillator with Hamiltonian $H_r=\hbar\omega_r a^\dagger a$, and directly couples to the DQD charge degree of freedom via its detuning.
This interaction can be described as $H_I=\hbar g_c(a^\dagger+a)\tau_z$, with $g_c$ the charge-photon coupling strength.
In the eigenbasis of $H_0$ this interaction acquires off-diagonal elements, which facilitates a spin-photon coupling $g_s\le g_c$ mediated by the charge states of the DQD (see \supmat\ Sec.~\ref{sup:DQDHamiltonianmodel}).
Near spin-photon resonance, the eigenenergies of the system form the Jaynes-Cummings ladder depicted in \figref{fig1:Samkh2018_comparison}(d) (see \supmat\ Sec.~\ref{supsec:JCmodel})~\cite{jaynes1963comparison}.

In the experiment, transitions between the system eigenstates are probed by measuring the transmission of a weak probe signal at frequency $f_{\text{probe}}=\omega_{\text{probe}}/2\pi$.
This coherent probe is described by a time-dependent driving term
\begin{equation} \label{eq:probeHamiltonian}
    V(t) = i\hbar\sqrt{\kappa_1}(e^{-i\omega_{\text{probe}} t} a_{\text{in,1}} a^\dagger -  e^{i\omega_{\text{probe}} t}a_{\text{in,1}}^* a),
\end{equation}
where $\kappa_1$ is the coupling strength between the resonator and the probe signal of coherent amplitude $a_{\text{in,1}}$. The probe power is related to this amplitude through $P_{\text{probe}} = \lambda\hbar\omega_{\text{probe}}\abs{a_{\text{in,1}}}^2$, where $\lambda$ accounts for extra losses in the probe signal delivery line \cite{gambetta2006qubit}.
For the two-tone simulations presented later in this work, a similar driving term $W(t)$ is added to describe a DQD pump tone with coherent amplitude $b_{\text{in}}$ (see \supmat\ Sec.~\ref{supsec:TwoToneIOT}).

Having described the system and its Hamiltonian, we now examine different classes of transitions that could match the spectrum in \figref{fig1:Samkh2018_comparison}(a).
The vacuum-Rabi-split modes correspond to the $\ket{\downarrow,0}\leftrightarrow\ket{1\pm}$ transitions (red) in the Jaynes-Cummings ladder [\figref{fig1:Samkh2018_comparison}(d)].
We find that the observed additional feature in the upper part of the gap closely matches the frequency of the $\ket{1+}\leftrightarrow\ket{2+}$ transition in the ladder, while the $\ket{1-}\leftrightarrow\ket{2-}$ transition frequency lies in the lower part of the gap, where no additional features are visible in the Samkharadze \etal\cite{samkharadze2018strong} experiment [\figref{fig1:Samkh2018_comparison}(a)].
Together, these transition frequencies (orange) form an eyelike shape in the middle of the gap.
Transitions involving higher states in the ladder, i.e., $\ket{m\pm}\leftrightarrow\ket{(m+1)\pm}$ for $m\ge2$ (transparent orange), move progressively closer to the middle of the spectrum for higher $m$.
Eventually they converge to a straight crossing of the modes that corresponds to the classical limit \cite{fink2010quantum}.
Circuit QED experiments with transmon qubits have reported observations of features corresponding to these excited-state transitions \cite{fink2008climbing}, as well as features corresponding to multiphoton transitions from the ground state to higher excited states in the ladder \cite{bishop2009nonlinear}. 
These multiphoton transitions form a fanlike structure in the spectrum (purple) and are not observed in the data from Samkharadze \etal\cite{samkharadze2018strong}.
Later in this work, we present new experiments with a different device that confirm the transition labeling.

To understand the relative visibility of these transitions, we now turn to an input-output description of the system. 
We first find the steady-state density matrix of the driven system from the Lindblad master equation
\begin{align} \label{eq2:masterequation}
    \begin{split}
        \frac{d\rho}{dt} = &-\frac{i}{\hbar}[H+V(t),\rho] + \gamma_1 \mathcal{D}[\widetilde\tau_-](\rho) + \frac{\gamma_\phi}{2} \mathcal{D}[\widetilde\tau_z](\rho)\\
        &+ (n_{\text{th}}+1)\kappa_r\mathcal{D}[a](\rho) + n_{\text{th}}\kappa_r\mathcal{D}[a^\dagger](\rho),
    \end{split}
\end{align}
with Lindblad dissipator $\mathcal{D}[A](\rho) = A\rho A^\dagger - \frac{1}{2}\{A^\dagger A,\rho\}$.
Charge relaxation (rate $\gamma_1$) and charge dephasing (rate $\gamma_\phi$) are described with Pauli operators $\widetilde\tau_-$ and $\widetilde\tau_z$ in the hybridized eigenbasis of charge states $\ket{\pm}$ \cite{benito2017input,benito2019optimized}.
The resonator linewidth $\kappa_r=\kappa_1+\kappa_2+\kappa_{\text{int}}$ consists of losses from coupling to the input-output lines ($\kappa_1=\kappa_2$) and internal losses ($\kappa_{\text{int}}$).
\figref{fig2:schematic_model} gives an overview of this model.
Spin decoherence due to nuclear spins in $^{28}$Si is much weaker than the decoherence caused by charge noise that couples in through spin-charge hybridization \cite{samkharadze2018strong,mi2018coherent}, and is therefore not included in this work.

The appearance of the feature inside the vacuum Rabi splitting requires a sufficient population of the excited states in the ladder, specifically the $\ket{1+}$ state.
For probe frequencies within the gap, excitation to these states by the coherent probe signal is suppressed, but could be caused by several other mechanisms.
Here, we empirically model incoherent excitations by coupling the resonator to a boson bath at temperature $T$ with a thermal occupation $n_{\text{th}} = 1/\left[\exp(\hbar\omega_r/k_{\text{B}} T)-1\right]$ \cite{fink2010quantum}.
For $g_c=0$, this will result in a thermal resonator state with temperature $T$, while for $g_c\neq 0$, this will lead to a finite population of excited DQD-resonator states. 
However, other mechanisms, like charge or spin excitation (or thermalization) effects can also populate the $\ket{1+}$ state, and can therefore produce similar signatures in the spectrum.
These mechanisms could not be differentiated here (see \supmat\ Sec.~\ref{supsec:thermalspin} for an example of a thermal spin model).

To find the steady state of the system, we first apply a multilevel rotating wave approximation (RWA) to get a time-independent master equation. 
We then truncate the resonator Hilbert space and transform all operators into the Liouville space \cite{manzano2020short,dzhioev2011super,harbola2008superoperator}, to arrive at a matrix-vector equation that can be numerically solved to find the steady-state density operator $\rho_S$ (see \supmat\ Sec.~\ref{supsec:IOTnumsol}).
The resonator transmission is then given by
\begin{equation}
    S_{21} = \frac{a_{\text{out,2}}}{a_{\text{in,1}}} = \frac{\langle \sqrt{\kappa_2}a \rangle}{a_{\text{in,1}}} = \frac{\sqrt{\kappa_2}\ \text{Tr}(a\rho_S)}{a_{\text{in,1}}}.
\end{equation}

The results from the experiment by Samkharadze \etal\cite{samkharadze2018strong} are well reproduced by simulations using this theoretical framework, as shown in \figref{fig1:Samkh2018_comparison}(c).
To obtain good agreement, we first determine the Hamiltonian parameters by matching the calculated Jaynes-Cummings transition frequencies to the experimental data.
We then manually adjust the bath temperature $T$, the probe amplitude $a_{\text{in,1}}$, and the charge decoherence rates $\gamma_1,\gamma_\phi$ to match the relative visibility of transitions in the spectrum.
The reason to proceed like this is mainly that the model has a large number of parameters that are underconstrained when fit to a single spectrum. 
Obtaining an automated fit would require simultaneously fitting to multiple heterogeneous datasets.
Alternatively, independent measurements can be used to determine certain parameters (more details in \supmat\ Sec.~\ref{supsec:simulationparams}).
 At $T=\SI{200}{\milli\kelvin}$, a finite population of excited Jaynes-Cummings states makes higher transitions in the ladder (orange) visible in the spectrum.
In this case, this leads to the appearance of a feature inside the vacuum Rabi splitting, which corresponds predominantly to the $\ket{1+}\leftrightarrow\ket{2+}$ transition.
Furthermore, the finite probe signal amplitude $a_{\text{in,1}} =\SI{1000}{\hertz}^{1/2}$ makes the main branches appear less bright near the top and bottom of the spectrum compared to the standard input-output simulation [\figref{fig1:Samkh2018_comparison}(b)].
This broadening of the spinlike transitions away from spin-photon resonance is also observed in the experiment [\figref{fig1:Samkh2018_comparison}(a)] and results from the finite population of higher-photon-number states generated by the probe signal.
Specifically, the simulated average photon number reaches $\langle a^\dagger a\rangle=\text{Tr}(a^\dagger a \rho_S)=0.31$ for $B_r=\SI{99}{\milli\tesla}$, which effectively broadens the spinlike transitions due to photon-number-dependent dispersive shifts \cite{harvey2022coherent}.
Increasing $a_{\text{in,1}}$ further in the simulations leads to a reduced vacuum Rabi splitting and the appearance of multiphoton transitions \cite{bishop2009nonlinear} in the spectrum.
However, these effects are not observed in the results from Samkharadze \etal\cite{samkharadze2018strong}, since the probe power was kept low in their experiment.
Finally, the charge decoherence rates $\gamma_1,\gamma_\phi$ make transitions in the lower part of the spectrum (involving $\ket{m-}$ states) more or less visible compared to features in the upper part (involving $\ket{m+}$ states) depending on their strength.
This effect is also observed in the experiment [\figref{fig1:Samkh2018_comparison}(a)] and simulations using the standard input-output theory [\figref{fig1:Samkh2018_comparison}(b)].
It is caused by an asymmetric admixture with the charge degree of freedom (i.e., the photonlike transition has less charge component below spin-photon resonance than above spin-photon resonance).
To match the relative visibility of upper and lower features in the experimental data, the simulations [\figref{fig1:Samkh2018_comparison}(b) and \figref{fig1:Samkh2018_comparison}(c)] use charge decoherence rates $\gamma_1/2\pi=\SI{20}{\mega\hertz}$ and $\gamma_\phi/2\pi=\SI{200}{\mega\hertz}$.

\begin{figure}[t]
    \centering
    \includegraphics{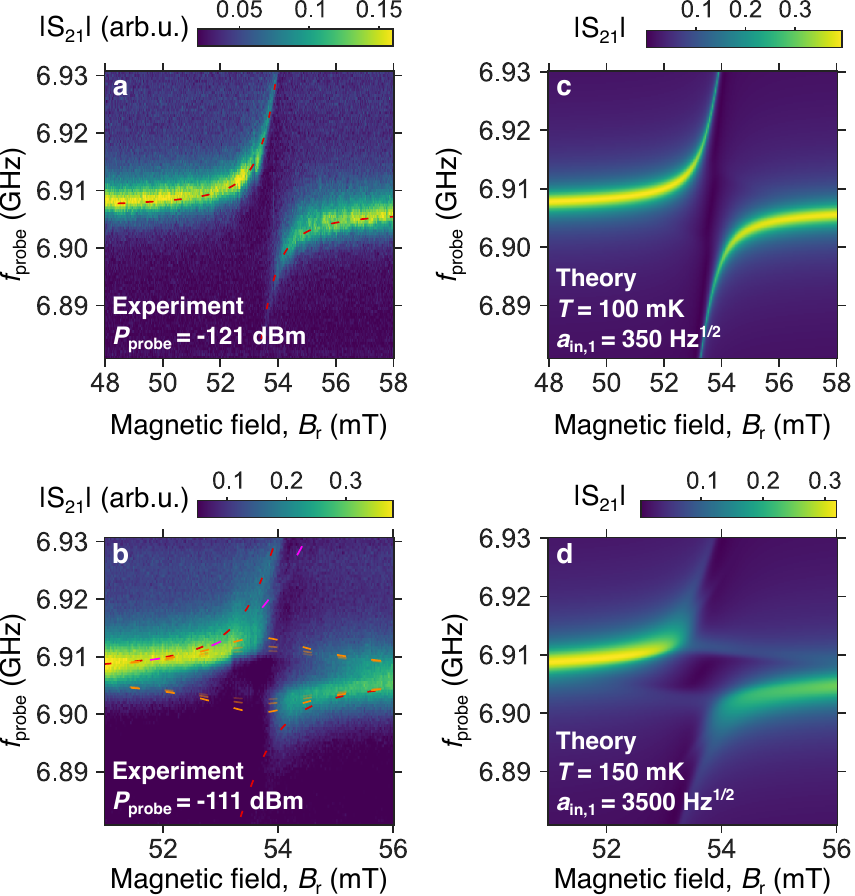}
    \caption{(a),(b) Resonator transmission spectra taken using the new device from Ref.~\cite{harvey2022coherent} at low (a,c) and high (b,d) probe power with relevant transition frequencies in the Jaynes-Cummings ladder, see \figref{fig1:Samkh2018_comparison}(d). Here, ``low'' and ``high'' probe power refers to the simulated average photon numbers $\langle a^\dagger a\rangle<0.1$ and $\langle a^\dagger a\rangle>1$ that are reached away from spin-photon resonance.  (c),(d) Simulated spectra using the theory presented in this work and the parameters in Supplemental Table~\ref{suptable:simparameters}.}
    \label{fig3:newexperiment_overview}
 \end{figure}

We now describe a new set of experiments in which we intentionally probe the transitions of the Jaynes-Cummings ladder described above.
The device, experimental setup, and data acquisition are described in detail in Ref.~\cite{harvey2022coherent} and were designed to realize resonator-mediated spin-spin interactions.
Here, we only use one of the DQDs (DQD2 in the nomenclature of Ref.~\cite{harvey2022coherent}), allowing its spin to interact with the resonator photons, while the other remains decoupled.
This system achieves a charge-photon coupling strength of $g_c/2\pi = \SI{192}{\mega\hertz}$ and is operated at a DQD tunnel coupling of $2t_c/h =\SI{12.0}{\giga\hertz}$ for these experiments, resulting in an effective spin-photon coupling strength of $g_s/2\pi \approx \SI{16}{\mega\hertz}$.
Since the bare resonator linewidth is $\kappa_r/2\pi = \SI{2.5}{\mega\hertz}$ and the spin linewidth is $\gamma_s/2\pi\le\SI{6}{\mega\hertz}$, the strong spin-photon coupling regime is achieved.
For a weak probe signal, the measured transmission spectrum in \figref{fig3:newexperiment_overview}(a) shows a simple avoided crossing of the main modes, while additional features are hardly visible.
The small dent in the upper branch around $B_r=\SI{53.2}{\milli\tesla}$ is believed to be an accidental crossing with a defect (two-level system).
When the probe power is increased, see \figref{fig3:newexperiment_overview}(b), features corresponding to both the $\ket{m+}\leftrightarrow\ket{(m+1)+}$ and $\ket{m-}\leftrightarrow\ket{(m+1)-}$ transitions become visible to form an eyelike shape in the spectrum (orange lines). 
Additionally, a faint feature appears near the upper branch that corresponds to the $\ket{\downarrow,0}\leftrightarrow\ket{2+}$ transition involving two-photon processes (purple line) \cite{bishop2009nonlinear}.

These results are well predicted by simulations using the theory developed for this work and shown in \figref{fig3:newexperiment_overview}(c) and \figref{fig3:newexperiment_overview}(d). 
To obtain good agreement, we employ the same manual fitting procedure as before and vary both the probe amplitude $a_{\text{in,1}}$ and the bath temperature $T$ between the low-power [\figref{fig3:newexperiment_overview}(c)] and high-power [\figref{fig3:newexperiment_overview}(d)] simulations.
The increase in $a_{\text{in,1}}$ leads to a fading of the branches near the top and the bottom of the spectrum, a reduced vacuum Rabi splitting, and the appearance of the $\ket{\downarrow,0}\leftrightarrow\ket{2+}$ transition in the simulated spectrum.
Interestingly, the high-power simulation uses an increased $T$ compared to the low-power simulation.
This increase in $T$ is needed to get agreement in the visibility of the eyelike feature, and might suggest a connection between the probe power and the effective temperature of the system.

\begin{figure*}[t]
    \centering
    \includegraphics{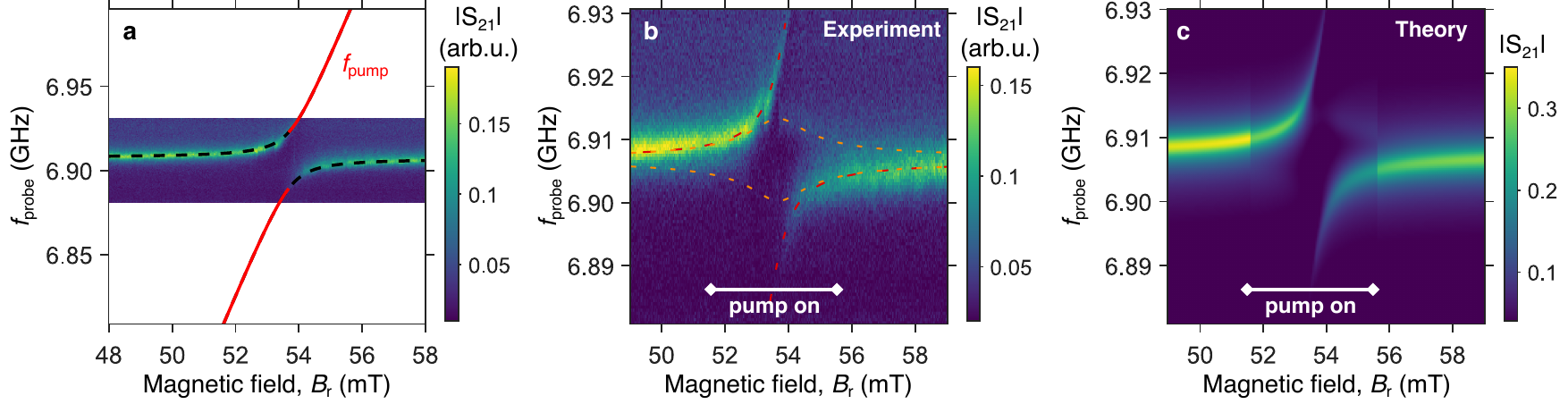}
    \caption{Two-tone spectroscopy scheme. (a) The $\ket{\downarrow,0}\leftrightarrow\ket{1\pm}$ transition frequencies are fitted to the spectrum at low probe power.
    The frequency of the additional pump tone (red line) is set to the $\ket{\downarrow,0}\leftrightarrow\ket{1-}$ transition (lower branch) for magnetic fields $\SI{51.60}{\milli\tesla}\le B_r\le \SI{53.65}{\milli\tesla}$ and to the $\ket{\downarrow,0}\leftrightarrow\ket{1+}$ transition (upper branch) for $\SI{53.65}{\milli\tesla}< B_r\le \SI{55.60}{\milli\tesla}$. (b)~Measured transmission spectrum with transition frequencies in the Jaynes-Cummings ladder, see \figref{fig1:Samkh2018_comparison}(d). (c) Simulated spectrum using the two-tone input-output model in \supmat\ Sec.~\ref{supsec:TwoToneIOT} for thermal bath temperature $T=\SI{200}{\milli\kelvin}$, pure charge dephasing rate $\gamma_\phi/2\pi=\SI{120}{\mega\hertz}$, and other parameters in Supplemental Table~\ref{suptable:simparameters}.} 
    \label{fig4:twotone_eyefeature}
\end{figure*}

Next, we reveal the eyelike transitions (orange) using a pump tone [see \figref{fig4:twotone_eyefeature}(a)] to generate population of the excited states \cite{fink2008climbing}.
This pump tone increases the steady-state occupation of the $\ket{1\pm}$ states, such that features corresponding to the $\ket{1\pm}\leftrightarrow\ket{2\pm}$ transitions become more apparent. 
The measured spectrum in \figref{fig4:twotone_eyefeature}(b) indeed reveals a feature corresponding to the $\ket{1+}\leftrightarrow\ket{2+}$ transition, while the feature corresponding to the $\ket{1-}\leftrightarrow\ket{2-}$ transition remains faint.
Using this pump-plus-weak-probe scheme, the extra Jaynes-Cummings transition appears in a more targeted way than in the previous strong-probe scheme of \figref{fig3:newexperiment_overview}(b) and \figref{fig3:newexperiment_overview}(d).

To model this two-tone experiment, a second driving term $W(t)$ that couples to the DQD detuning is added to the master equation in \eqnref{eq2:masterequation}. 
Since the Hamiltonian then contains terms rotating at two different frequencies, the RWA fails to eliminate the time dependence in the master equation and we can no longer find a steady-state solution as before.
To circumvent this issue, we assume the probe signal is weak and calculate the resonator transmission in the linear response regime (see \supmat\ Sec.~\ref{supsec:TwoToneIOT}).
The simulated spectrum using this approach in \figref{fig4:twotone_eyefeature}(c) shows good qualitative agreement with the measured data. 
The sharp changes of visibility in the simulated spectrum appear due to the switching on and off of the pump tone, and are also observed in the experiment to some degree.

In summary, we have observed additional transitions in the vacuum Rabi splitting spectrum of spin circuit QED devices.
We have identified these transitions as involving higher excited states in the Jaynes-Cummings ladder, thereby also explaining previously reported observations.
The visibility of these transitions was enhanced by increasing the probe power and by using a pump-and-probe scheme.
We found the experimental data to be in agreement with simulations using an input-output framework based on a steady-state solution of a Lindblad master equation. 
Improvements in the coupling-to-decoherence ratio (cooperativity) enable more distinct observations of these transitions, allowing one to probe higher transitions in the Jaynes-Cummings ladder.
In that regard, the new experiments presented here are a witness of the improvements in cooperativity in this spin-photon system.
In the future, selective driving of these transitions could prove useful for photon preparation and measurement schemes~\cite{johnson2010quantum,albert2018performance}.
Finally, the input-output framework presented in this work can be straightforwardly extended to accurately describe resonator-mediated interactions between two spins, which pave the way to a scalable spin qubit architecture \cite{borjans2020resonant,harvey2022coherent,vandersypen2017interfacing}.

\section*{Acknowledgements}
The authors thank G.\ Zheng for his contributions to setting up the experiment, L.~P.\ Kouwenhoven and his team for access to the NbTiN film deposition, F.\ Alanis Carrasco for assistance with sample fabrication, and other members of the spin qubit team at QuTech for useful discussions.

\textbf{Funding}\ \ \
This research was undertaken thanks in part to funding from the European Research Council (ERC Synergy Quantum Computer Lab), and the Dutch Ministry for Economic Affairs through the allowance for Topconsortia for Knowledge and Innovation (TKI).

\textbf{Author contributions}\ \ \
P.H.-C. and T.B. identified the relevant transitions from the Hamiltonian model.
T.B. and M.R. developed the theoretical framework.
P.H.-C., J.D., and T.B. performed the electrical cryogenic measurements.
P.H.-C. fabricated the device.
A.S. contributed to sample fabrication. A.S. grew the heterostructure with G.S.’s supervision.
T.B., P.H.-C., J.D., M.R., and L.M.K.V. analyzed the results.
T.B. wrote the manuscript with input from all co-authors.
P.H.-C. and L.M.K.V. supervised the project.

\textbf{Data availability}\ \ \
The data and simulation scripts used in this paper are archived online at  \url{https://dx.doi.org/10.4121/19336748}. 

\footnotesize
%
%


\makeatletter
\renewcommand\thesection{\mbox{S\arabic{section}}}
\makeatother
\setcounter{section}{0}     

\newcounter{supfigure} \setcounter{supfigure}{0} 
\makeatletter
\renewcommand\thefigure{\mbox{S\arabic{supfigure}}}
\makeatother

\newcounter{suptable} \setcounter{suptable}{0} 
\makeatletter
\renewcommand\thetable{\mbox{S\arabic{suptable}}}
\makeatother

\newenvironment{supfigure}[1][]{\begin{figure}[#1]\addtocounter{supfigure}{1}}{\end{figure}}
\newenvironment{supfigure*}[1][]{\begin{figure*}[#1]\addtocounter{supfigure}{1}}{\end{figure*}}
\newenvironment{suptable}[1][]{\begin{table}[#1]\addtocounter{suptable}{1}}{\end{table}}
\newenvironment{suptable*}[1][]{\begin{table*}[#1]\addtocounter{suptable}{1}}{\end{table*}}

\makeatletter
\renewcommand\theequation{S\arabic{equation}}
\renewcommand\theHequation{S\arabic{equation}} 
\makeatother
\setcounter{equation}{0}


\onecolumngrid 
\clearpage
{\centering
\large\textbf
{Supplementary information for: \\ \mytitle} \\ \rule{0pt}{12pt}
}
\twocolumngrid


\section{Input-output theory} \label{supsec:IOT}

\subsection{DQD eigenbasis} \label{sup:DQDHamiltonianmodel}
In earlier work, Benito \etal\cite{benito2017input} derived analytical expressions for the spin-charge hybridized eigenstates and eigenenergies of the DQD Hamiltonian $H_0$ (Eq.~(1) of the main text).
By first expressing $H_0$ in the product basis of antibonding and bonding orbitals $\ket{\pm}$ with spin $\uparrow,\downarrow$, they found the DQD energy levels to be
\begin{align}
    \begin{split}
        E_{3,0}=\pm\frac{1}{2}\Bigg[\bigg(\Omega+g_e\mu_B &\sqrt{B_z^2+B_x^2\sin^2\theta}\ \bigg)^2\\
        &+(g_e\mu_B B_x)^2\cos^2\theta\Bigg]^{1/2},
    \end{split}
\end{align}
\begin{align}
    \begin{split}
        E_{2,1}=\pm\frac{1}{2}\Bigg[\bigg(\Omega-g_e\mu_B&\sqrt{B_z^2+B_x^2\sin^2\theta}\ \bigg)^2\\
        &+(g_e\mu_B B_x)^2\cos^2\theta\Bigg]^{1/2},
    \end{split}
\end{align}
where $\Omega = \sqrt{\epsilon^2 + 4t_c^2}$ is the charge qubit energy splitting and $\theta=\arctan(\epsilon/2t_c)$ is the orbital angle.
The energetically close states $\ket{1}$ and $\ket{2}$ experience a strong hybridization and are given by
\begin{equation}\label{eq:DQD1}
    \ket{1} = \cos\frac{\Phi}{2}\ket{-,\uparrow}+\sin\frac{\Phi}{2}\ket{+,\downarrow},
\end{equation}
\begin{equation}\label{eq:DQD2}
    \ket{2} = \sin\frac{\Phi}{2}\ket{-,\uparrow}-\cos\frac{\Phi}{2}\ket{+,\downarrow},
\end{equation}
where $\Phi = \arctan\frac{g_e\mu_B B_x \cos\theta}{\Omega-g_e\mu_B B_z}$ is the spin-orbit mixing angle. The remaining eigenstates are approximated as
\begin{equation}\label{eq:DQD0}
    \ket{0}\approx \ket{-,\downarrow},
\end{equation}
\begin{equation}\label{eq:DQD3}
    \ket{3}\approx\ket{+,\uparrow}.
\end{equation}
Operators acting on the DQD states that appear in the master equation in Eq.~(3) are conveniently expressed in this eigenbasis of $H_0$.
Specifically, dipole ``raising'' and ``lowering'' operators are introduced that describe the interaction between the DQD and electric fields applied to its gates. 
The dipole raising operator $d_{+}$ describes an excitation of the DQD by incident fields and is written in the eigenbasis $\{\ket{0},\ket{1},\ket{2},\ket{3}\}$ (Eqs.~\eqref{eq:DQD1}~--~\eqref{eq:DQD3}) as
\begin{equation}
    d_+ = \begin{bmatrix} 0 & 0 & 0 & 0 \\
                        d_{01} & 0 & 0 & 0 \\
                        d_{02} & 0 & 0 & 0 \\
                        0 & d_{13} & d_{23} & 0 \end{bmatrix},
\end{equation}
while a deexcitation is described by $d_-=d_+^\dagger$, with matrix elements 
\begin{equation}
    d_{01} = d_{23} \approx -\cos\theta\sin\frac{\Phi}{2},
\end{equation}
\begin{equation}
    d_{02} = -d_{13} \approx \cos\theta\cos\frac{\Phi}{2}.
\end{equation}
The experiments in this work operate in the regime where the $\ket{0}\leftrightarrow\ket{1}$ transition of the DQD is predominantly spin-like ($\cos\Phi>0$), such that the effective spin-photon coupling strength becomes $g_s=g_c\abs{d_{01}}$ \cite{benito2017input}.
Furthermore, the DQD decoherence operators describing charge relaxation ($\widetilde\tau_-$) and pure charge dephasing ($\widetilde\tau_z$) can be expressed in this basis as
\begin{align}
    \begin{split}
        \widetilde\tau_-= &\begin{bmatrix} 0 & \sin(\Phi/2) & -\cos(\Phi/2) & 0\\
                                  0 & 0 & 0 & \cos(\Phi/2)\\  
                                  0 & 0 & 0 & \sin(\Phi/2)\\
                                  0 & 0 & 0 & 0\end{bmatrix},\\
        \widetilde\tau_z &= \begin{bmatrix} -1 & 0 & 0 & 0\\
                                  0 & -\cos\Phi & -\sin\Phi & 0\\  
                                  0 & -\sin\Phi & \cos\Phi & 0\\
                                  0 & 0 & 0 & 1\end{bmatrix}.
    \end{split}
\end{align}

\subsection{Magnetic field model} \label{supsec:micromagnetmodel}
The DQD Hamiltonian in Eq. (1) of the main text includes the homogenous magnetic field component $B_z$ and the interdot magnetic field difference $2B_x$. 
The longitudinal magnetic field difference $2b_z$ is engineered to be small, yielding a transverse spin-photon coupling \cite{beaudoin2016coupling}.
Transmission measurements as a function of DQD detuning and applied magnetic field strength provide estimates for $b_z$ that are below $\SI{1}{\milli\tesla}$ for the devices considered in this work \cite{croot2020flopping}. 
Such a longitudinal magnetic field difference leads to a small asymmetry of the DQD energy levels as a function of detuning and a correction of the spin energy which can be expressed as $-g_e \mu_B b_z \epsilon/ \Omega$ \cite{benito2019electric}.
Since $b_z\ll B_z,B_x$ and the experiments in this work are performed at $\eps=\SI{0}{\micro\electronvolt}$, this residual longitudinal effect is small and not necessary to include. 
It can, however, lead to unnecessary spin dephasing via charge noise.

The local magnetic fields at the DQD, $B_z$ and $B_x$, have contributions from both the external and micromagnet fields.
To accurately model the experiments, we express these local fields in terms of the applied field $\Vec{B}_{\text{ext}}$ using the micromagnet model from Ref.~\cite{harvey2022coherent}.
This non-trivial model was developed to capture the full dependence of the DQD energy levels on the magnitude and direction of the applied magnetic field.
Here, we use it to compute the homogenous magnetic field component $B_z$ from the applied magnetic field strength $B_r$, while the interdot magnetic field difference $2B_x$ is assumed to be constant.
To this end, the magnetic field experienced by an electron in the left (right) dot is defined as $\Vec{B}_{L(R)}$ \cite{beaudoin2016coupling}. 
The average magnetic field $\Vec{B}$ in the DQD can then be expressed as
\begin{equation}
    \Vec{B} = (\Vec{B}_L + \Vec{B}_R)/2 = \Vec{B}_{\text{ext}} + \Vec{B}_{\upmu\text{m}}.
\end{equation}
Here the micromagnet average field $\Vec{B}_{\upmu\text{m}}$ in the DQD is modeled with the empirical formula
\begin{equation}
    \Vec{B}_{\upmu\text{m}} = \left(B_{\upmu\text{m}0} + \chi_{\upmu\text{m}} (\Vec{B}_{\text{ext}} - \Vec{B}_{\text{ext,0}})\cdot \hat{\boldsymbol{u}}_{\upmu\text{m}}\right)\hat{\boldsymbol{u}}_{\upmu\text{m}},
\end{equation}
where $\hat{\boldsymbol{u}}_{\upmu\text{m}}$ is the micromagnet unit vector, $\chi_{\upmu\text{m}}$ is the micromagnet susceptibility, and $B_{\upmu\text{m}}$ and $\Vec{B}_{\text{ext,0}}$ are constant offsets. 
Vectors in this model are conveniently expressed in spherical coordinates. For example, the external magnetic field is expressed as
\begin{equation}
    \Vec{B}_{\text{ext}} = (B_r,\theta = -\ang{90},\phi),
\end{equation}
with $B_r$ the applied magnetic field magnitude and $\phi$ the polar angle. 
The micromagnet unit vector $\hat{\boldsymbol{u}}_{\upmu\text{m}}$ follows from the geometry of the device.
The $z$ axis is then chosen to point along the average magnetic field in the DQD such that $B_z = \abs{\Vec{B}}$.
The micromagnet parameters $\chi_{\upmu\text{m}}, B_{\upmu\text{m}0}, \Vec{B}_{\text{ext,0}}, B_x$ are determined by fitting the resulting transition frequencies to the measured resonator transmission spectra as in Ref.~\cite{harvey2022coherent}. 
We find this model to be sufficient to capture the magnetic field dependence of the spin Zeeman energies over the range of interest.

\renewcommand{\arraystretch}{1.2}
\begin{suptable*}[t]
\begin{tabular}{lccc}
\cline{1-4}
\textbf{Parameter} & \textbf{Symbol} & \textbf{Samkharadze \etal\cite{samkharadze2018strong}} & \textbf{This work}  \\ \cline{1-4}
 Bare resonator frequency  &    $\omega_r/2\pi$            & \SI{6.051}{\giga\hertz} & \SI{6.916}{\giga\hertz}\\
 Bare resonator linewidth  &    $\kappa_r/2\pi$              & \SI{2.7}{\mega\hertz} & \SI{2.5}{\mega\hertz}\\
 Internal resonator decay rate &$\kappa_{\text{int}}/2\pi$ & \SI{1.46}{\mega\hertz} & $\approx\SI{1.5}{\mega\hertz}$\\
 Maximal resonator photon number (single-tone) & $N$ & 10 & 15 \\
 Maximal resonator photon number (two-tone) & $N$ & - & 10 \\
 Charge-photon coupling strength  &    $g_c/2\pi$          & \SI{200}{\mega\hertz} & \SI{192}{\mega\hertz}\\
 Micromagnet unit vector        & $\hat{\boldsymbol{u}}_{\upmu\text{m}}$ & $(1,\ang{90},\ang{0})$ & $(1,\ang{270},\ang{15})$\\
 Micromagnet susceptibility  & $\chi_{\upmu\text{m}}$ & 0.23 & 0.63\\
 Initial micromagnet field   &    $B_{\upmu\text{m}0}$             & \SI{90}{\milli\tesla} & \SI{147.5}{\milli\tesla}\\
 External magnetic field polar angle & $\phi$ & $\ang{0}$ & $\ang{10.6}$\\
 External magnetic field offset  &    $\Vec{B}_{\text{ext,0}}$     & $\begin{pmatrix}\SI{-20.5}{\milli\tesla}\le B_{r0}\le \SI{-13.5}{\milli\tesla}\\ \ang{-90}\\ \ang{0}\end{pmatrix}$ & $\begin{pmatrix} B_{r0}= \SI{-20}{\milli\tesla}\\ \ang{-90}\\ \ang{10.6}\end{pmatrix}$\\
 DQD magnetic field difference   &  $2B_x$    &   \SI{25}{\milli\tesla} & \SI{60}{\milli\tesla} \\
 DQD detuning & $\epsilon$  & \SI{0}{\micro\electronvolt} & \SI{0}{\micro\electronvolt}\\
 DQD tunnel coupling & $2t_c/h$  & 7.8 - \SI{14.6}{\giga\hertz} & \SI{12.0}{\giga\hertz}\\
 Charge relaxation rate         &  $\gamma_1/2\pi$  & \SI{20}{\mega\hertz}  &  \SI{1}{\mega\hertz} \\
 Pure charge dephasing rate (single-tone)     &  $\gamma_\phi/2\pi$  & \SI{200}{\mega\hertz} & \SI{10}{\mega\hertz} \\
 Pure charge dephasing rate (two-tone)     &  $\gamma_\phi/2\pi$  & - & \SI{120}{\mega\hertz} \\
 Pump tone amplitude (two-tone)     &  $b_{\text{in}}$  & - & $\SI{5000}{\hertz}^{1/2}$ \\
 Pump tone coupling strength (two-tone)     &  $\delta/2\pi$  & - & \SI{1}{\mega\hertz} \\ \cline{1-4}
\end{tabular}
\caption{Parameters used to simulate spin circuit QED experiments with the devices from Samkharadze \etal~\cite{samkharadze2018strong} and this work (see \refref{harvey2022coherent} for details on this device). The simulation parameters are discussed in more detail in \secref{supsec:simulationparams}.}
\label{suptable:simparameters}
\end{suptable*}

\subsection{Jaynes-Cummings model} \label{supsec:JCmodel}
In the experiments described here, the charge qubit energy splitting $2t_c$ is detuned from both the spin and photon energies.
The dispersive charge-photon interaction leads to a shift of the photon transition, i.e., $\omega_r\rightarrow\omega_r-\chi_c$, where $\chi_c$ includes shifts from both co-rotating and counter-rotating (Bloch-Siegert shift) terms \cite{kohler2018dispersive,zeuch2020exact} and is given by
\begin{equation}
    \chi_c \approx \frac{(g_c)^2}{\Omega/\hbar-\omega_r} + \frac{(g_c)^2}{\Omega/\hbar+\omega_r}.
\end{equation}
When we use this shifted resonator frequency and apply a rotating wave approximation (RWA), we can reduce the system to a spin-photon Jaynes-Cummings model \cite{jaynes1963comparison}.  
Near spin-photon resonance, the excited eigenstates of the coupled system can then be expressed as the $m$-excitation states
\begin{align}
    \begin{split}
        \ket{m+} &= \sin(\alpha_n/2) \ket{\downarrow,m} + \cos(\alpha_n/2) \ket{\uparrow,m-1},\\
        \ket{m-} &= \cos(\alpha_n/2) \ket{\downarrow,m} - \sin(\alpha_n/2) \ket{\uparrow,m-1},\\
    \end{split}
\end{align}
with spin-photon mixing angle $\alpha_n = \frac{1}{2}\arctan(2g_s\sqrt{n+1}/\Delta)$, where $\Delta=E_1-E_0-\omega_r$. 
Here $\ket{\uparrow,n}$ ($\ket{\downarrow,n}$) denotes the state with the electron in the spin $\uparrow$ ($\downarrow$) state and $n$ photons in the resonator. 
The corresponding energies are
\begin{equation}
    E_{m\pm} = m\hbar\omega_r \pm \frac{\hbar}{2}\sqrt{\Delta^2+4g_s^2m},
\end{equation}
and $E_g=-\hbar\Delta/2$ for the ground state $\ket{g}=\ket{\downarrow,0}$ \cite{blais2004cavity}.
These dressed-state energies form the Jaynes-Cummings ladder, which is depicted in Fig.~1d of the main text for the case of $\Delta=0$.

\begin{supfigure*}
    \centering
    \includegraphics{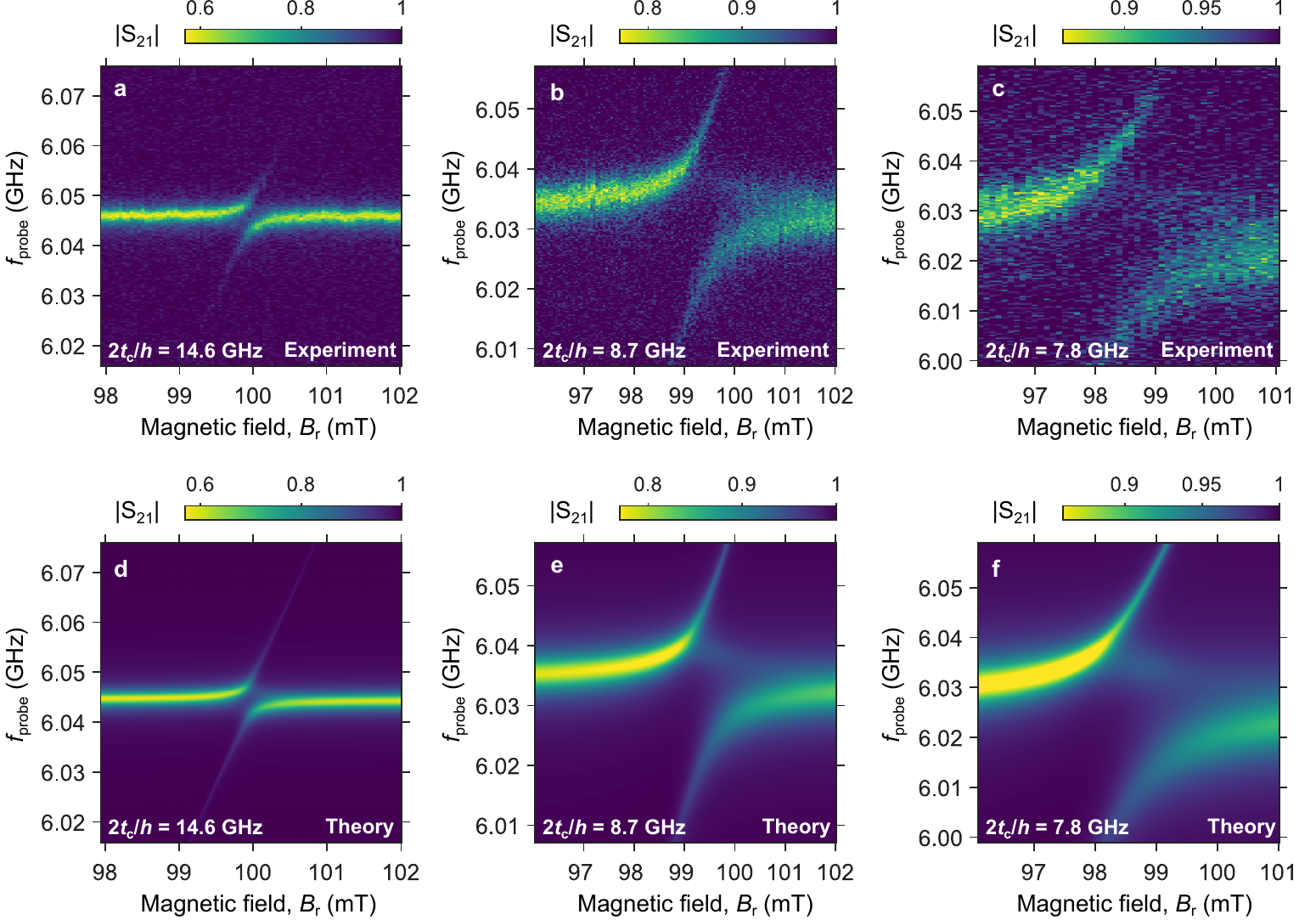}
    \caption{Tunnel coupling dependence of the vacuum Rabi splitting from Samkharadze \etal\cite{samkharadze2018strong}. \textbf{(a-c)} Experimental data for the indicated values of the DQD tunnel coupling. \textbf{(d-f)} Simulated spectra for probe amplitude $a_{\text{in,1}} =\SI{1000}{\hertz}^{1/2}$, thermal bath temperature $T=\SI{200}{\milli\kelvin}$ and other parameters in Table \ref{suptable:simparameters}, including up to $N=10$ photons in the resonator Hilbert space.}
    \label{figS1:Science_tc_dependence}
\end{supfigure*}

\subsection{Standard input-output theory} \label{supsec:standardIOT}
The input-output theory for spin circuit QED developed by Benito \etal\cite{benito2017input} has become a standard in the field.
Their theory is based on a steady-state solution of Quantum Langevin equations (QLEs) for the DQD ($\sigma_{ij}=\ket{i}\bra{j}$, with $\ket{i}$ the DQD eigenstates) and resonator ($a^{(\dagger)}$) operators.
The complex resonator transmission $S_{21}$ is derived within a RWA by 
introducing DQD transition susceptibilities $\chi_{ij}$.
When the DQD is in its ground state, the relevant susceptibilities are expressed as
\begin{equation} \label{eq:chi01}
    \chi_{01} = \frac{g_c\cos\theta\sin(\Phi/2)}{\delta_1-i\gamma_{\text{eff}}^{(2)}},
\end{equation}
\begin{equation}\label{eq:chi02}
    \chi_{02} = \frac{-g_c\cos\theta\cos(\Phi/2)}{\delta_2-i\gamma_{\text{eff}}^{(1)}},
\end{equation}
with detunings $\delta_n=E_n-E_0-\omega_{\text{probe}}$ and effective decoherence rates $\gamma_{\text{eff}}^{(n)}=(\gamma_1/2+\gamma_\phi)[\delta_2\sin^2(\Phi/2)+\delta_1\cos^2(\Phi/2)]/\delta_n$ \cite{benito2017input}.
The resonator transmission is written in terms of these susceptibilities as
\begin{equation}\label{eq:S21standardIOT}
    S_{21} = \frac{-i\sqrt{\kappa_1\kappa_2}}{\omega_r-\omega_{\text{probe}}-i\kappa/2+g_c(\chi_{01}d_{01}+\chi_{02}d_{02})}.
\end{equation}

Fig.~1b shows the spectrum that is predicted by this standard input-output theory for the experiment by Samkharadze \etal\cite{samkharadze2018strong}.
When compared to the experimental data in Fig.~1a and the simulation using the theory presented in this work in Fig.~1c, we see that the theory by Benito \etal\cite{benito2017input} does not reproduce the additional feature within the gap.
This is because the additional feature corresponds to transitions between entangled states in the Jaynes-Cummings ladder ($\ket{m\pm}$), which are not captured by the standard input-output theory since it assumes separable steady states, i.e., $\langle a\sigma_{ij}\rangle=\langle a\rangle\langle \sigma_{ij}\rangle$.
This treatment captures transitions from a separable state to entangled spin-photon states, most importantly the vacuum Rabi split $\ket{\downarrow,0}\leftrightarrow\ket{1\pm}$ transitions.
However, transitions where both the initial and the final states are entangled spin-photon states are not captured in this separable state ansatz.
Since the observed additional feature corresponds to transitions between entangled states in the Jaynes-Cummings ladder ($\ket{m\pm}$, orange transitions in Fig.~1d), it cannot be reproduced by this model (Fig.~1b).
As we have shown, the theoretical framework developed for this work does capture these effects.
However, it should be noted that it is significantly more computationally heavy compared to the standard input-output theory. 

Furthermore, the main branches of the vacuum Rabi splitting appear brighter near the top and bottom of the spectrum predicted by the standard input-output theory.
As was discussed in the main text, this fading of the main branches is a result of a finite probe power.
Since Eqs.~\eqref{eq:chi01}~--~\eqref{eq:S21standardIOT} do not include any dependence on probe strength ($a_{\text{in,1}}$), this effect is not captured by the standard input-output theory.

Finally, we note that Eqs.~\eqref{eq:chi01}~--~\eqref{eq:S21standardIOT} can be generalized to arbitrary DQD level occupations in a straightforward way.
A finite population of excited DQD states leads to a smaller vacuum Rabi splitting compared to the result when the DQD is in the ground state.
However, this generalization cannot reproduce the additional transitions reported in this work, since it still uses the separable state ansatz discussed above.

\subsection{Multi-level RWA} \label{supsec:multilevelRWA}
In order to arrive at a time-independent master equation, we move into a frame rotating with the probe frequency. The Hamiltonian in this frame then becomes $\Tilde{H} = UHU^\dagger + i\hbar\frac{dU}{dt}U^\dagger$, with
\begin{equation} \label{eq:RWAunitary}
    U = \exp(i\omega_{\text{probe}} t(a^\dagger a + \ket{1}\bra{1} + \ket{2}\bra{2} + 2\ket{3}\bra{3})).
\end{equation}
Carrying out this transformation and neglecting all fast-oscillating terms in a rotating wave approximation (RWA), the Hamiltonian terms become
\begin{equation}
    \Tilde{H}_0 = \begin{bmatrix} E_0 & 0 & 0 & 0\\
                                  0 & E_1-\hbar\omega_{\text{probe}} & 0 & 0\\  
                                  0 & 0 & E_2-\hbar\omega_{\text{probe}} & 0\\
                                  0 & 0 & 0 & E_3-2\hbar\omega_{\text{probe}}\end{bmatrix},
\end{equation}
\begin{equation}
    \Tilde{H}_r = \hbar(\omega_r-\omega_{\text{probe}})a^\dagger a,
\end{equation}
\begin{equation}
    \Tilde{H}_I = \hbar g_c \left(a^\dagger d_- + a d_+ \right),
\end{equation}
\begin{equation}
    \Tilde{V} = i\hbar\sqrt{\kappa_1}\left(a_{\text{in,1}} a^\dagger - a_{\text{in,1}}^* a\right).
\end{equation}
However, counter-rotating terms that are neglected in the RWA lead to significant shifts in the energy levels of the system. 
This results in a shift of the resonator frequency known as the Bloch-Siegert shift \cite{zeuch2020exact}.
We include this shift in our input-output model by substituting the resonator frequency, i.e., $\omega_r\rightarrow\omega_r-\chi^{\text{BS}}_c$, where the Bloch-Siegert shift is approximated by 
\begin{equation}
    \chi^{\text{BS}}_c \approx \frac{(g_c)^2}{\Omega/\hbar+\omega_r}.
\end{equation}

\begin{supfigure*}
    \centering
    \includegraphics{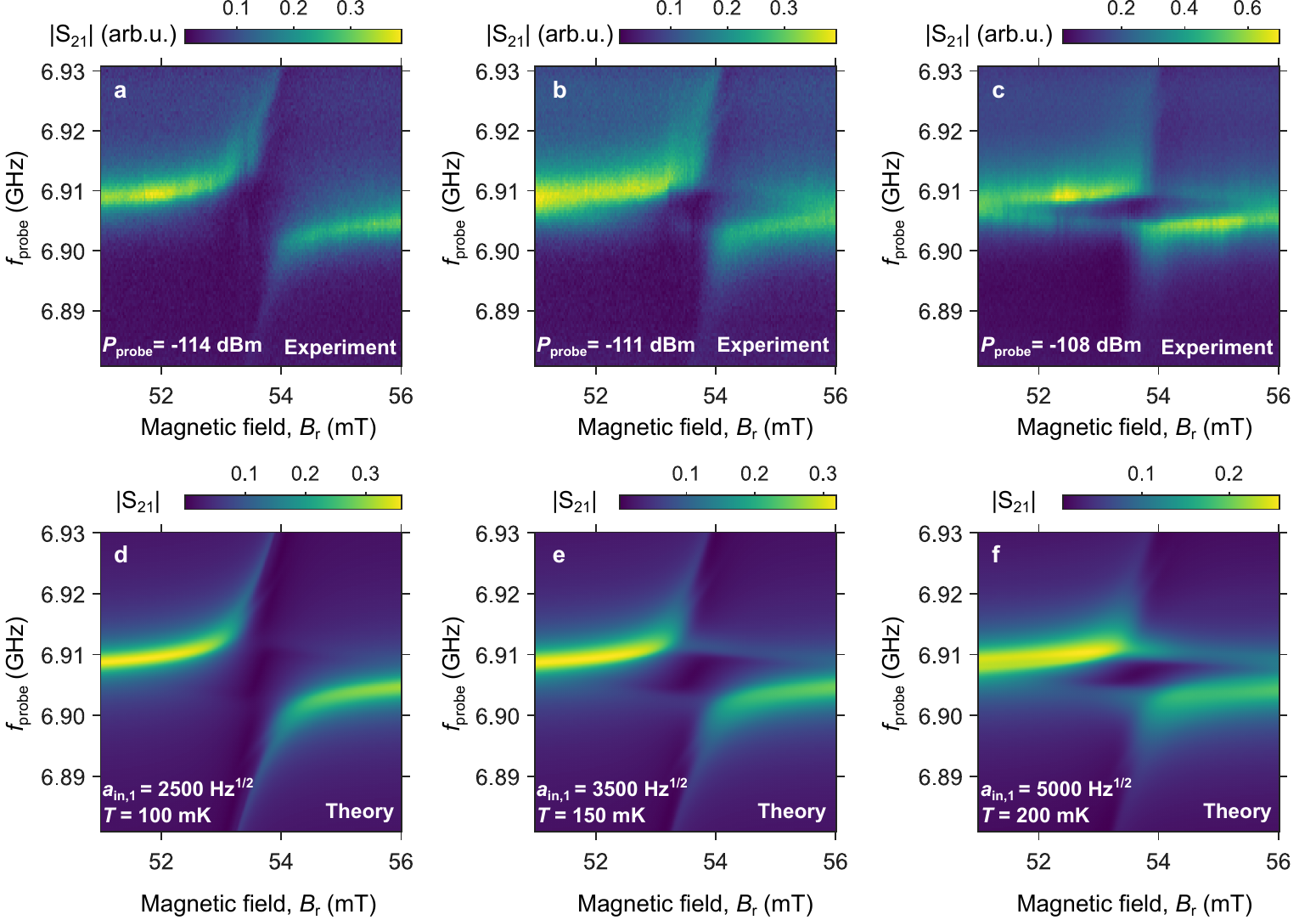}
    \caption{Probe power dependence of the vacuum Rabi splitting for the new device of Ref.~\cite{harvey2022coherent}. \textbf{(a-c)} Experimental data for the indicated values of the probe power. \textbf{(d-f)} Simulated spectra for the indicated values of the probe amplitude $a_{\text{in,1}}$ and thermal bath temperature $T$, and other parameters in Table \ref{suptable:simparameters}, including up to $N=15$ photons in the resonator Hilbert space.}
    \label{figS2:extended_probe_power_dependence}
\end{supfigure*}

\subsection{Numerical solution} \label{supsec:IOTnumsol}
To find a numerical solution of the master equation in the steady state, we truncate the Hilbert space of the resonator, such that it includes Fock states $\ket{n}$, where the resonator photon number $n$ ranges from 0 to $N$. The master equation can then be written in the product basis of resonator and DQD states $\{\ket{i}\otimes\ket{n}\}$. This leads to a matrix equation that can be solved for $\rho$, which now has dimension $M=4(N+1)$. To transform this to an easily solvable matrix-vector equation, we transform the operators to the so-called Liouville space \cite{manzano2020short,dzhioev2011super,harbola2008superoperator}. The dimension of this space is $M^2$ and in this representation, the density matrix becomes a vector $\ket{\rho}$,
\begin{equation}
    \rho = \sum_{n,m=0}^{M-1} \rho_{nm} \ket{n}\bra{m} \rightarrow \ket{\rho} = \sum_{n,m=0}^{M-1} \rho_{mn} \ket{n}\otimes\ket{\Tilde{m}},
\end{equation}
where $\ket{\Tilde{m}}$ is a state vector in the Hilbert space that is an identical copy of the original one.
In this representation, the master equation in the rotating frame can be written as
\begin{equation}
    \frac{d}{dt} \ket{\Tilde{\rho}} = \mathcal{L}\ket{\Tilde{\rho}} = \left(-\frac{i}{\hbar}(\mathcal{H}+\mathcal{V})+\mathcal{D}\right)\ket{\Tilde{\rho}},
\end{equation}
where $\ket{\Tilde{\rho}}$ is a vector that represents the density matrix in the rotating frame and Lindbladian $\mathcal{L}$ is now an $M\times M$ matrix, which consists of the Hamiltonian $\mathcal{H}$, driving term $\mathcal{V}$, and dissipator $\mathcal{D}$ in the Liouville space. 
The steady-state density matrix can then be found by solving $\mathcal{L}\ket{\Tilde{\rho}_S} = 0$.
To directly find a normalized solution, we replace the first row of $\mathcal{L}$ to impose the trace condition $\Tr(\rho)=1$, thereby converting the steady-state master equation to a problem of the form
\begin{equation}
    \mathcal{L}'\ket{\Tilde{\rho}_S} = y,
\end{equation}
where $y=(1,0,0,0,\ldots)^T$. This matrix-vector equation can then be solved using standard methods \cite{bishop2009nonlinear,bishop2010circuit}.
This method of calculating the steady-state density matrix was found to be faster than the direct computation and normalization of the Null space of $\mathcal{L}$.

\subsection{Two-tone spectroscopy} \label{supsec:TwoToneIOT}
In order to model two-tone spectroscopy experiments, we add a coherent driving field that couples to the DQD charge dipole. 
This pump tone uses the same coupling mechanism as the resonator field and we again use dipole operators $d_\pm$ to express the corresponding driving term as
\begin{equation} \label{eq:pumpHamiltonian} 
    W(t) = i\hbar\sqrt{\delta}(e^{-i\omega_{\text{pump}} t} b_{\text{in}} d_+ -  e^{i\omega_{\text{pump}} t}b_{\text{in}}^* d_-).
\end{equation}
Here $\omega_{\text{pump}}$ is the frequency of this pump tone, $b_{\text{in}}$ is the coherent amplitude, and $\delta$ is the coupling strength between the DQD charge dipole and this coherent drive. 
Analogously to \eqnref{eq:RWAunitary}, we now transform to a frame rotating with the pump frequency $\omega_{\text{pump}}$. 
The resulting master equation reads
\begin{equation} \label{eq:twotoneLME}
    \frac{d\Tilde{\rho}}{dt} = -\frac{i}{\hbar}[\Tilde{H}+\Tilde{W}+\Tilde{V}(t),\Tilde{\rho}] + \mathcal{D}(\Tilde{\rho}),
\end{equation}
where $\Tilde{W}$ and $\Tilde{V}(t)$ represent the Hamiltonian terms corresponding to the pump (\eqnref{eq:pumpHamiltonian}) and probe (Eq.~(2)) tone, respectively. 
Since the probe driving term is still time-dependent, this equation cannot be numerically solved in the same way as before. 
Instead, one has to resort to time-dependent simulations or the Floquet formalism. 
Here, we take a different approach and assume the probe signal to be weak, i.e., small $a_{\text{in,1}}$, which is often the case in circuit QED experiments. 
To calculate the steady-state resonator transmission in this regime, we first neglect the probe driving term, such that the master equation in the stationary limit becomes
\begin{equation}
    0 = -\frac{i}{\hbar}[\Tilde{H}+\Tilde{W},\Tilde{\rho}_S] + \mathcal{D}(\Tilde{\rho}_S).
\end{equation}
This equation can then be solved to find the steady-state density matrix $\Tilde{\rho}_S$ under coherent excitation from the pump tone. 

The transmission of the probe signal to the resonator is then calculated in the linear response regime \cite{landig2019virtual,kohler2018dispersive}.
To see how this works, we transform the master equation in \eqnref{eq:twotoneLME} to Fourier space and separate the commutator,
\begin{equation}
    i\omega\Tilde{\rho}_\omega = -\frac{i}{\hbar}[\Tilde{H}+\Tilde{W},\Tilde{\rho}_\omega] + \mathcal{D}(\Tilde{\rho}_\omega) -\frac{i}{\hbar}[\Tilde{V}(t),\Tilde{\rho}(t)]_\omega,
\end{equation}
where the subscript $\omega$ denotes the Fourier transform of the operators and we used the fact that the only time dependencies in the master equation are carried by $\Tilde{V}(t)$ and $\Tilde{\rho}(t)$.
The time dependence of the last commutator in this equation leads to a convolution in frequency space. 
Next, the probe drive $\Tilde{V}(t)$ is assumed to be a weak perturbation to the steady-state density matrix of the system $\Tilde{\rho}_S$ in the absence of a probe drive \cite{kohler2018dispersive}. 
Including the perturbation caused by the probe only to first order, the master equation is simplified to
\begin{equation}
    i\omega\Tilde{\rho}_\omega = -\frac{i}{\hbar}[\Tilde{H}+\Tilde{W},\Tilde{\rho}_\omega] + \mathcal{D}(\Tilde{\rho}_\omega) -\frac{i}{\hbar}[\Tilde{V}_\omega,\Tilde{\rho}_S].
\end{equation}
The Fourier transform of the probe drive Hamiltonian is given by
\begin{align}
	\begin{split}
		\Tilde{V}_\omega = i\hbar\sqrt{\kappa_1}(\delta(&\omega+\omega_{\text{probe}}-\omega_{\text{pump}})a_{\text{in,1}} a^\dagger \\
		-& \delta(\omega-\omega_{\text{probe}}+\omega_{\text{pump}}) a_{\text{in,1}}^* a),
	\end{split}
\end{align}
where $\delta(x)$ is the Dirac delta function. 
As was shown in \refref{kohler2018dispersive}, the effect of the second term in this equation can be neglected. 
The resonator response is then found by writing the master equation in the Liouville space, yielding
\begin{equation}
    i\omega \ket{\Tilde{\rho}_\omega} = \left(-\frac{i}{\hbar}(\mathcal{H}+\mathcal{W})+\mathcal{D}\right)\ket{\Tilde{\rho}_\omega} - \frac{i}{\hbar}\mathcal{V}_\omega\ket{\Tilde{\rho}_S}.
\end{equation}
The probe drive term in this space can be written as
\begin{align}
\begin{split}
    \mathcal{V}_\omega = i\hbar\sqrt{\kappa_1}a_{\text{in,1}}\left(a^\dagger\otimes\mathbb{I}_M - \mathbb{I}_M\otimes (a^\dagger)^T\right) \delta(\omega+\\ \omega_{\text{probe}}-\omega_{\text{pump}}).
\end{split}
\end{align}
The master equation can then be solved to find the perturbed density matrix
\begin{align}
    \begin{split}
        \ket{\Tilde{\rho}} = i\hbar\sqrt{\kappa_1}a_{\text{in,1}}(-\mathcal{H}-\mathcal{W}-i\hbar\mathcal{D} +\hbar(\omega_{\text{probe}}\\ -\omega_{\text{pump}})\mathbb{I}_{M^2})^{-1} \left(a^\dagger\otimes\mathbb{I}_M - \mathbb{I}_M\otimes (a^\dagger)^T\right)\ket{\Tilde{\rho}_S},    
    \end{split}
\end{align}
where the equation integrated over frequency space to move back to the time domain. 
The full resonator transmission in the linear response regime can finally be computed from the density matrix as
\begin{equation}
    S_{21} = \frac{a_{\text{out,2}}}{a_{\text{in,1}}} = \frac{\sqrt{\kappa_2}\langle a \rangle}{a_{\text{in,1}}}  = \frac{\sqrt{\kappa_2}}{a_{\text{in,1}}}\bra{\mathbb{I}_M} (a\otimes\mathbb{I}_M)\ket{\Tilde{\rho}}.
\end{equation}

\subsection{Determining the simulation parameters} \label{supsec:simulationparams}
The parameters that were used to obtain the simulated transmission spectra presented in this work are listed in Table \ref{suptable:simparameters}.
The bare resonator frequency and linewidths were measured with the DQDs away from zero detuning.
The remaining Hamiltonian parameters were determined by matching the calculated transition frequencies in the Jaynes-Cummings ladder to measured transmission spectra.
Finally, coherent drive parameters $a_{\text{in,1}}$, $b_{\text{in}}$, $\delta$, bath temperature $T$, and charge decoherence rates $\gamma_1,\gamma_\phi$ were manually adjusted to match the relative visibility of transitions in the spectrum.
The significant digits in the parameters give a qualitative sense of the degree of confidence in their precise value, while statistical error bars would not reflect our real degree of confidence.

The reason to proceed like this is mainly that the model has a large number of parameters that are underconstrained when fit to a single spectrum. 
Obtaining an automated fit would require simultaneously fitting to multiple heterogeneous data sets. 
Alternatively, independent measurements can be used to determine certain parameters. 
For example, resonator parameters can be determined with high confidence far away from zero detuning and used in the subsequent determination of new parameters. 
Conversely, some parameters can change between datasets because they are tuneable or unstable. 
Performing an actual simultaneous fit to the full datasets was not feasible because of these requirements. 
Most importantly, constrained fits to subsets of data would yield statistical error bars that do not reflect our real confidence in the parameter.
Secondly, the simulations take a significant amount of time to run on a simulation server. 
This is especially the case for the simulations with a high probe power (high $a_{\text{in,1}}$), because they require including states with up to $N=15$ photons. 
Alternatively, one could pick certain linecuts in the spectrum, which would have to be properly weighted to capture the confidence in them. Although this might be possible, it is still somewhat arbitrary. 
We decided not to pursue this, since the goal of this work is to explain the physics involved in the appearance of these features. 
Full quantitative agreement would need further refinements to the methodology.

We note that for the data from Samkharadze \etal\cite{samkharadze2018strong}, slight modifications of $B_{\text{r0}}$ between spectra were needed to account for hysteresis effects in the micromagnets. Furthermore, as was stated in the main text, the device used for the new experiments in this work uses a transmission-style resonator coupling, leading to a peak in the transmission when the system is probed on resonance. 
In contrast to the hanger-style coupling in Samkharadze \etal\cite{samkharadze2018strong}, this coupling provides no information on the maximal transmission. 
Since the precise amount of losses and amplification between the system and instruments at room temperature is not known, the measured $\abs{S_{21}}$ has arbitrary units and we are unable to extract the internal resonator decay rate $\kappa_{\text{int}}$. 
We therefore assume $\kappa_{\text{int}}\approx\SI{1.5}{\mega\hertz}$ based on knowledge of previous devices with similar resonator designs.

To reproduce the observed asymmetry between upper and lower features in the two-tone spectrum in Fig.~4, a higher charge dephasing rate $\gamma_\phi/2\pi=\SI{120}{\mega\hertz}$ is used than in the single-tone experiments in Fig.~3 ($\gamma_\phi/2\pi=\SI{10}{\mega\hertz}$).
This seems to suggest that driving the DQD also increases the decoherence in the system.
The drive is applied through a gate line that has an on-chip microwave filter \cite{harvey2020chip,harvey2022coherent}, and we have seen that this generates significant cross-talk between the two DQDs.
It is possible that this also contributes to extra dephasing of the charge or spin degrees of freedom, in which case changing the cut-off frequency of the filter for this drive line would help mitigate the issue.

\section{Extended data}
In this section, we present more data revealing additional features in the resonator transmission spectrum. 
The vacuum Rabi splittings at several values of the DQD tunnel coupling, measured in Samkharadze \etal\cite{samkharadze2018strong}, are reproduced here in \figref{figS1:Science_tc_dependence}a-c.
At large $t_c$, the hybridization of spin and charge remains small, as does the resulting spin-photon coupling.
When $2t_c/h$ is brought closer to the resonator frequency (hence reduced in this case), the spin-charge admixing becomes larger, leading to a larger vacuum Rabi splitting in the spectrum.
Moreover, both the DQD spin qubit and the resonator become more vulnerable to charge-induced decoherence, leading to larger linewidths, as well as a larger difference in visibility between upper and lower features in the spectrum. 
Simulations using the input-output theory presented in this work are shown in \figref{figS1:Science_tc_dependence}d-f and are in agreement with these observations.

\begin{supfigure}[t]
    \centering
    \includegraphics{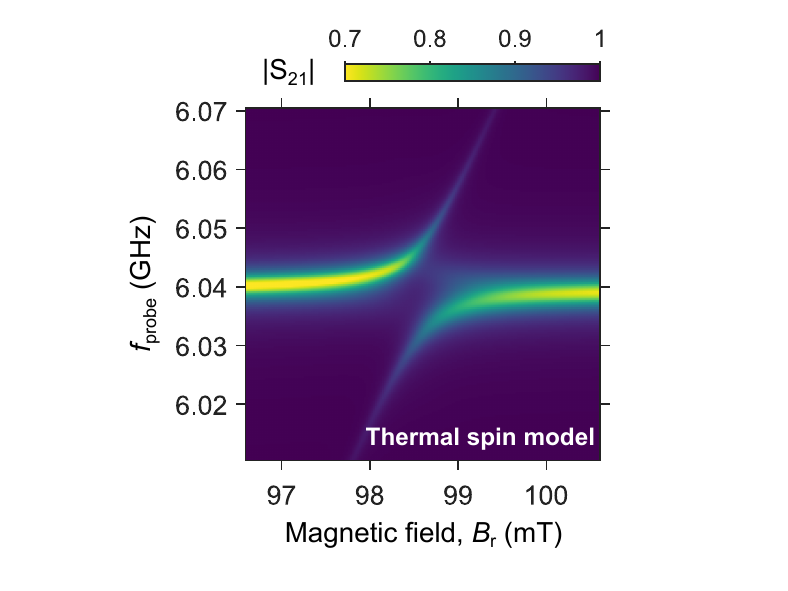}
    \caption{Simulation of the experiment by Samkharadze \etal\cite{samkharadze2018strong} for $2t_c/h=\SI{10.4}{\giga\hertz}$, including a thermal bath at temperature $T_{\text{spin}}=\SI{300}{\milli\kelvin}$ that is coupled to the DQD spin with coupling strength $\gamma_s/2\pi=\SI{1}{\mega\hertz}$, as described by master equation \eqnref{eq:masterequation_thermalspin}.
    The probe amplitude $a_{\text{in,1}} =\SI{1000}{\hertz}^{1/2}$ and other parameters listed in Table \ref{suptable:simparameters} are the same as for the simulation in Fig.~1c, where a thermal bath was coupled to the resonator instead of the DQD spin.}
    \label{figS3:ThermalSpinmodel_Science2018}
\end{supfigure}

Secondly, \figref{figS2:extended_probe_power_dependence}a-c show the transmission spectra over a range of probe powers in our new experiment.
Starting from the low-power spectrum in \figref{figS2:extended_probe_power_dependence}a, increasing the probe power leads to a fading of the main branches near the top and bottom of the spectrum, a reduced vacuum Rabi splitting, and the appearance of transitions between excited states (eye-like shape) and a two-photon transition, see Fig.~3 of the main text.
The simulated spectra are in agreement with these results.
To obtain this level of agreement, we vary both the probe amplitude $a_{\text{in,1}}$ and the temperature of the thermal bath $T$.
Since the probe powers in the measured spectra are separated by \SI{3}{\decibel m}, the corresponding scaling factor in $a_{\text{in,1}}$ can be found using the relation in the main text to be $\sqrt{2}$.
Using this scaling leads to reasonable agreement in the fading of the main branches, the reduction of the vacuum Rabi splitting, and the appearance of the two-photon transition near the upper branch.
A faint feature corresponding to the $\ket{\downarrow,0}\leftrightarrow\ket{2-}$ two-photon transition appears near the lower branch in the simulated spectra, while it is not as clear in the measured spectra.
In addition to this scaling of the coherent probe amplitude, an increase in the population of excited Jaynes-Cummings states is needed to reproduce the observed power dependence of the eye-like shape in the spectrum (corresponding to orange transitions in Fig.~1d).
This increase is modeled by empirically increasing the temperature of the thermal bath. 
While it is possible that this is indeed caused by heating of the system, other sources of incoherent excitations could also play a role (see \secref{supsec:thermalspin}).

\newpage\section{Incoherent excitation mechanisms} \label{supsec:thermalspin}
As discussed in the main text, the appearance of transitions between excited Jaynes-Cummings states requires a finite population of the excited states in the ladder.
In the master equation Eq.~(3), we have modeled incoherent excitations by coupling the resonator to a thermal bath.
However, other mechanisms can also populate excited states in the ladder, leading to similar signatures in the resonator transmission spectrum.
As an example, we show that coupling a thermal bath to the DQD spin degree of freedom reproduces the results of Samkharadze \etal\cite{samkharadze2018strong} in Fig.~1 of the main text.
To this end, we solve a different master equation
\begin{align} \label{eq:masterequation_thermalspin}
    \begin{split}
        \frac{d\rho}{dt} = &-\frac{i}{\hbar}[H+V(t),\rho] + \kappa_r\mathcal{D}[a](\rho) + \gamma_1 \mathcal{D}[\widetilde\tau_-](\rho)  \\
        &+ \frac{\gamma_\phi}{2} \mathcal{D}[\widetilde\tau_z](\rho) + \gamma_s \mathcal{D}[\sqrt{\boldsymbol{n_{\text{th}}}+1}\odot\sigma_-](\rho)\\
        &+ \gamma_s\mathcal{D}[\sqrt{\boldsymbol{n_{\text{th}}}}\odot\sigma_+](\rho),
    \end{split}
\end{align}
which includes photon losses ($\kappa_r$), charge relaxation ($\gamma_1$), pure charge dephasing ($\gamma_\phi$), and coupling of the spin to a thermal bath with coupling strength $\gamma_s$.
Here $\odot$ denotes the element-wise matrix product and $\boldsymbol{n_{\text{th}}}$ is a matrix containing the thermal bath occupations at the DQD transition frequencies, i.e., $\boldsymbol{n_{\text{th}}}^{ij} = 1/\left(\exp(\abs{E_i-E_j}/k_BT)-1\right)$.
The spin relaxation operator is written in the $H_0$ eigenbasis as 
\begin{equation}
    \sigma_- = \begin{bmatrix} 0 & \cos(\Phi/2) & \sin(\Phi/2) & 0\\
                                  0 & 0 & 0 & \sin(\Phi/2)\\  
                                  0 & 0 & 0 & -\cos(\Phi/2)\\
                                  0 & 0 & 0 & 0\end{bmatrix},
\end{equation}
and the spin excitation operator is $\sigma_+=\sigma_-^\dagger$.
The steady-state solution of this master equation is found using the methods described in \secref{supsec:IOT}.

\figref{figS3:ThermalSpinmodel_Science2018} shows a simulation of the experiment by Samkharadze \etal\cite{samkharadze2018strong} using this thermal spin model.
The resulting spectrum again reveals a feature corresponding to the $\ket{1+}\leftrightarrow\ket{2+}$ transition in the Jaynes-Cummings ladder and looks similar to the simulation in Fig.~1c of the main text, where a thermal bath was coupled to the resonator instead of the spin.
These two mechanisms both populate excited states in the Jaynes-Cummings ladder and cannot be differentiated using the experiments reported here.
In future work, accurate measurements of the DQD level occupations and the photon number (e.g., using the method in \refref{harvey2022coherent}) for $g_c=0$ could reveal which mechanism is more present in this system. \clearpage
\end{document}